\documentclass[aps,prl,twocolumn,showpacs]{revtex4}

\usepackage{dcolumn}
\usepackage{amssymb}
\usepackage{graphicx}

\begin{document}

\newcommand \be {\begin{equation}}
\newcommand \ee {\end{equation}}
\newcommand \bea {\begin{eqnarray}}
\newcommand \eea {\end{eqnarray}}
\newcommand \nn {\nonumber}

\title{Temperature in nonequilibrium systems with conserved energy}
\author{Eric Bertin,$^{1}$ Olivier Dauchot,$^2$ and Michel Droz$^1$}
\affiliation{$^1$Department of Theoretical Physics, University of Geneva, CH-1211 Geneva 4, Switzerland\\
$^2$SPEC, CEA Saclay, F-91191 Gif-sur-Yvette Cedex, France}
\date{\today}

\begin{abstract}
We study a class of nonequilibrium lattice models which describe local redistributions of a globally conserved energy. A particular subclass can be solved analytically, allowing to define a temperature $T_{th}$ along the same lines as in the equilibrium microcanonical ensemble. The fluctuation-dissipation relation is explicitely found to be linear, but its slope differs from the inverse temperature $T_{th}^{-1}$. A numerical renormalization group procedure suggests that, at a coarse-grained level, all models behave similarly,
leading to a two-parameter description of their macroscopic properties.
\end{abstract}


\maketitle

Understanding the behaviour of nonequilibrium systems with a large number of degrees of freedom remains one of the major goals for statistical physics. Many attemps have been made to describe such systems in terms of a limited set of macroscopic parameters, like in the equilibrium case \cite{Jou}. In the context of glasses (i.e.~materials with huge relaxation times), an effective temperature --first introduced in a phenomenological way \cite{Tool,Naraya,Moynihan}-- has been defined from linear fluctuation-dissipation relations (FDR), on the basis of mean-field spin-glass models \cite{CuKu}, and was shown to satisfy the basic properties needed to define a temperature \cite{CuKuPe}. Since then, this FDR has been tested numerically \cite{Parisi,Kob,Barrat00,Sciortino,Berthier,Zamponi,Ritort} and experimentally \cite{Israeloff,Ciliberto,Ocio,Danna,Abou} in many realistic glassy systems. Still, no clear consensus concerning the good definition of temperature has emerged since the measured FDR are not always line

Apart from glassy materials, other classes of systems having a finite relaxation time (granular gases, non-hamiltonian spin models, etc.) can be considered as out-of-equilibrium in the sense that their dynamics does not fulfill detailed balance (DB). Their steady state cannot be described in general in the framework of equilibrium statistical physics, even though they can share some quantitative properties with equilibrium systems --e.g.~critical behavior \cite{Grinstein,Tauber}. A few attemps \cite{Miller,Baldassarri,Barrat03,Chate} have been made to define an effective temperature from Jaynes' maximum entropy condition \cite{Jaynes} or from FDR, but the interpretation of the resulting temperatures remains to be clarified.

Since the breaking of DB plays a key role in nonequilibrium systems, it is worth distinguishing several forms of DB. What is often refered to as DB in the literature is a canonical version:
\be
W(\beta|\alpha)\, e^{-E_{\alpha}/T} = W(\alpha|\beta)\, e^{-E_{\beta}/T}
\ee
where $W(\beta|\alpha)$ is the transition rate from state $\alpha$ to state $\beta$. This form is just a simple way to enforce canonical equilibrium when defining a stochastic model, hence its usefulness for numerical simulations. Still, one could wish to define stochastic models in a microcanonical situation. In this case, the stochastic dynamics should obvioulsy conserve the energy, and one can also assume a microcanonical DB relation $W(\beta|\alpha) = W(\alpha|\beta)$. Actually this form of DB --also called microreversibility-- is not just a useful recipe but can be given a fundamental interpretation in an equilibrium context, as it is associated to the time-reversal symmetry of the underlying hamiltonian dynamics.

Turning to nonequilibrium systems, one expects on general grounds that the dynamics breaks the time-reversal symmetry due to the presence of fluxes or dissipation. So it may be of interest to study the most simple nonequilibrium stochastic systems which are defined by relaxing only the microreversibility condition, replacing it by a more general DB relation $W(\beta|\alpha) f_{\alpha} = W(\alpha|\beta) f_{\beta}$, while preserving the energy conservation.

In this Letter, we study a class $\cal C$ of nonequilibrium lattice models which describe local redistributions of a globally conserved energy. A particular subclass ${\cal C}_s$, defined later on, satisfying DB (but not always microreversibility) can be solved analytically, allowing to define a temperature $T_{th}$ along the same lines as in the equilibrium microcanonical ensemble. Moreover, we derive explicitely the FDR and show that it is linear, with a slope different from the inverse temperature $T_{th}^{-1}$, thus questioning the relevance of FDR to define temperature in non glassy out-of-equilibrium systems. In addition, a functional renormalization group procedure implemented numerically suggests that any model from  class $\cal C$ behaves macroscopically like a member of the subclass ${\cal C}_s$, indicating that DB is restored on a coarse-grained level.

Our models are defined as follows. A real variable $-\infty < x_i < \infty$ is attached to each site $i$ of a $d$-dimensional hypercubic lattice with $N$ sites. The (pseudo-)energy $E= \frac{1}{2} \sum_{i=1}^N x_i^2$ is conserved by the dynamics defined by the following stochastic rules.
At each time step, a link $(i,j)$ is randomly picked up on the lattice, and a random number $q \in [0,1]$ is drawn from a symmetric distribution $\psi(q)$. 
Then the variables $x_i$ and $x_j$ are updated as:
\be \label{redist}
x_i' = \pm \sqrt{q\, (x_i^2+x_j^2)}, \quad
x_j' = \pm \sqrt{(1-q) \,(x_i^2+x_j^2)}
\ee
The sign $\pm$ is randomly chosen with equal probability. 
The different models in $\cal C$ are distinguished by $\psi(q)$.
These dynamical rules can be formulated in terms of a master equation with transitions rates $W(\{x_i'\}|\{x_i\})$.
It is generaly hopeless to find the stationary solution of a master equation unless the DB condition is fulfilled. 
Let us define the subclass ${\cal C}_s$, consisting of the models for which
$\psi(q)$ is chosen to be a beta distribution:
\be
\psi(q) = \frac{\Gamma(2\eta)}{\Gamma(\eta)^2} \, q^{\eta-1} (1-q)^{\eta-1}, \qquad \eta >0
\ee
It can be shown that in this case, DB holds, namely:
\be \label{eq-DB}
W(\{x_i'\}|\{x_i\}) \prod_{i=1}^N |x_i|^{2\eta-1}= W(\{x_i\}|\{x_i'\}) \prod_{i=1}^N |x_i'|^{2\eta-1}
\ee
Technical details will be reported elsewhere \cite{long}.

Thus microreversibility is recovered only for $\eta=\frac{1}{2}$ --a case similar to $\eta=1/2$ has been studied in \cite{Presutti}. The steady-state distribution is readily obtained from Eq.~(\ref{eq-DB}) by taking into account the energy conservation and normalizing the resulting distribution:
\be \label{dist-stat}
P_{st}(\{x_i\}) = \frac{1}{Z_N(E)} \, \prod_{i=1}^N |x_i|^{2\eta-1} \; \delta \left( \sum_{i=1}^N \frac{x_i^2}{2} - E \right)
\ee
with $Z_N(E) = K_N E^{\eta N -1}$ and $K_N = 2^{\eta N} \Gamma(\eta)^N/\Gamma(\eta N)$. For $\eta \ne \frac{1}{2}$, this distribution is clearly non uniform over the states of given energy; one can then expect important differences with equilibrium systems.

In the framework of equilibrium microcanonical ensemble, a well define
prescription exists for introducing temperature.
One considers an isolated system (with constant energy) and divides it into two subsystems. Temperature is introduced as a thermodynamic parameter which takes equal values in both subsystems. If moreover the value of this parameter is independent of the choice of the partition, the parameter can be said to characterize the whole system. 

We consider a partition into subsystems ${\cal S}_1$ and ${\cal S}_2$ which can exchange energy while keeping $E_1+E_2=E$ fixed. The energies $E_1$ and $E_2$ are fluctuating, but in the limit of large subsystems, the mean value of $E_k$ can be identified with its most probable value $E_k^*$. Generalizing the equilibrium procedure, the relevant quantity to compute is then the conditional probability $P(E_1|E)$ that the subsystem ${\cal S}_1$ has energy $E_1$ given that the total energy is $E$. Using Eq.~(\ref{dist-stat}), one finds:
\be
P(E_1|E) = \frac{Z_{N_1}(E_1)\, Z_{N_2}(E-E_1)}{Z_N(E)}
\ee
In the usual equilibrium microcanonical ensemble, $Z_N(E)$ reduces to the phase-space area $\Omega_N(E)$ of the hypersurface with energy $E$.
The most probable energy $E_1^*$ is found from $\partial \ln P/\partial {E_1} \vert_{E_1^*} = 0$, which gives:
\be \label{mut-equil}
\frac{\partial \ln Z_{N_1}}{\partial E_1} \Big\vert_{E_1^*} = \frac{\partial \ln Z_{N_2}}{\partial E_2} \Big\vert_{E_2^*}
\ee
This allows to define a temperature $T_{th}^k$ for each subsystem $k$ through (we set $k_B=1$):
\be
\frac{1}{T_{th}^k} \equiv \frac{\partial \ln Z_{N_k}}{\partial E_k} \Big\vert_{E_k^*}
\ee
Thus Eq.~(\ref{mut-equil}) implies $T_{th}^1 = T_{th}^2$. It can be checked that the common value does not depend on the partition chosen \cite{long}, so that this temperature can be safely said to characterize the whole system.

From the expression of $Z_N(E)$, one finds $T_{th} = \varepsilon/\eta$, where 
$\varepsilon = E/N$ is the energy density. Also, considering subsystem ${\cal S}_1$ as very small with respect to ${\cal S}_2$, but still macroscopic, one can derive in a similar way the `canonical' probability distribution:
\be \label{dist-can}
P_{can}(\{x_i\}) = \frac{1}{Z_{N_1}^{can}} \prod_{i=1}^{N_1} |x_i|^{2\eta-1} \, \exp \left( -\frac{\sum_{i=1}^{N_1} x_i^2}{2\, T_{th}} \right)
\ee

Another way to introduce a temperature in non-equilibrium systems is to consider generalized FDR. This approach has received considerable attention since it has been given a precise interpretation in the context of glasses \cite{CuKuPe}. Still, its applicability for non glassy out-of-equilibrium system remains to be clarified, and can be tested within the present model. To this aim, an external field $h$ must be introduced to allow for the definition of a response function. A natural way to include an external field is to add to the energy $E$ a perturbing term $-h\sum_i x_i$; one can rewrite the energy $E_h = \frac{1}{2} \sum_{i=1}^N (x_i-h)^2$ up to an irrelevant additive constant. Note that $E_h$ has the same form as $E$ in terms of the variables $v_i \equiv x_i-h$. The dynamics of the $v_i$'s is then defined in the same way as for the $x_i$'s in the absence of field, which is consistent with the equilibrium procedure. One then recovers for $v_i$ the canonical distribution Eq.~(\ref{dist-can}).

To compute the response function, we assume that the system, subjected to a field, is in steady state for $t<0$. At time $t=0$, the field is switched off. The response is defined for $t>0$ by $\chi(t) = \partial <N^{-1} \sum_i x_i>/\partial h \vert_{h=0}$. From the canonical distribution, the following FDR is derived:
\be \label{eq-FDR}
\chi(t) = \frac{1}{T_{th}}\, \langle x_i(t) x_i(0) \rangle_{h=0}
- (2\eta-1) \left< \frac{x_i(t)}{x_i(0)} \right>_{h=0}
\ee
Although Eq.~(\ref{eq-FDR}) does not lead at first sight to a linear relation between $\chi(t)$ and $\langle x_i(t) x_i(0) \rangle$, some simplifications actually occur. Indeed, it can be seen that correlation functions of the form $\langle x_i(t)^n x_i(0)^m \rangle$ with odd integers $n$ and $m$ are all proportional to the `hopping correlation function' $\Phi(t) = \langle N^{-1} \sum_{i=1}^N \phi_i(t) \rangle$, with $\phi_i(t) = 1$ if $x_i(t)=x_i(0)$ and $\phi_i(t)=0$ otherwise. More specifically, $\langle x_i(t)^n x_i(0)^m \rangle = \langle x_i(0)^{n+m} \rangle \, \Phi(t)$ \footnote{As long as no transition involving site $i$ occurred, $x_i(t)^n x_i(0)^m$ remains equal to $x_i(0)^n x_i(0)^m$. But when site $i$ is updated, the new value $x_i'$ becomes completely decorrelated from $x_i(0)$ due to the random sign of $x_i'$.}.

As a result, one has $\langle x_i(t) x_i(0) \rangle = 2\varepsilon \Phi(t)$, and $\langle x_i(t) / x_i(0) \rangle = \Phi(t)$, so that the FDR~(\ref{eq-FDR}) can be rewritten:
\be
\chi(t) = \frac{1}{2\varepsilon}\, \langle x_i(t) x_i(0) \rangle
\ee
yielding a fluctuation-dissipation (FD) temperature $T_{FD} =  2\varepsilon$, which is different from $T_{th}=\varepsilon/\eta$ (except in the case $\eta=\frac{1}{2}$ for which microreversibility is recovered).

This leads us to the question: which of the two temperatures $T_{th}$ and $T_{FD}$ is more relevant from a physical point of view? Actually, it could be argued that both temperatures may be equivalent up to a redefinition of the temperature unit: if $T_{th}$ takes the same value in two subsystems, so does $T_{FD}$. Still, this conclusion only remains valid as long as $\eta$ takes the same value throughout the system. It is then natural to test a more general dynamics. Interestingly, DB still holds if one introduces on each link a different distribution $\psi_{ij}(q)$:
\be
\psi_{ij}(q) = \frac{\Gamma(\eta_i+\eta_j)}{\Gamma(\eta_i) \, \Gamma(\eta_j)} \, q^{\eta_i-1} (1-q)^{\eta_j-1}
\ee
where $\eta_i$ can take a different value on each site $i$; $q$ refers to site $i$ and $1-q$ to site $j$ as in Eq.~(\ref{redist}). Then the microcanonical distribution $P_{st}(\{x_i\})$ takes the same form as in Eq.~(\ref{dist-stat}), simply replacing $\prod_i |x_i|^{2\eta-1}$ by $\prod_i |x_i|^{2\eta_i-1}$. Following the same lines as above, $P(E_1|E)$ is easily computed and leads to equal values of the temperature $T_{th}^k$ in both subsystems, with $T_{th} = \varepsilon_k / \langle \eta \rangle_k$ ($\langle \eta \rangle_k$ denotes a spatial average of $\eta_i$ over subsystem $k$). On the contrary, the FDR formally keeps the same form as previously, and the FD temperature remains related to the energy density through $T_{FD} = 2 \varepsilon$. Choosing $\{\eta_i\}$ such that $\langle \eta \rangle_1 \neq \langle \eta \rangle_2$, the equality $T_{th}^1 = T_{th}^2$ implies $\varepsilon_1 \neq \varepsilon_2$. The equipartition of energy breaks down, which in turn leads to $T^1_{FD} \neq T^2_{FD}$. Note that si

Thus the two temperatures $T_{th}$ and $T_{FD}$ are not equivalent up to a change of temperature unit, but differ qualitatively since $T_{FD}$ does not necessarily take the same value in two systems which can freely exchange energy. Still, if $T_{th}$ was not known, one could have tried to argue that $T_{FD}$ is the correct temperature, in a spirit similar to the procedure invoked in \cite{CuKuPe} for glassy systems. Indeed, connecting a new site acting as a thermometer to the existing system, one may identify its average energy with $\frac{1}{2}T$, as done also to define a granular temperature \cite{Haff}. Interestingly, this yields precisely the same result as $T_{FD}$, i.e.~$T=2\varepsilon$ (assuming a uniform $\eta$). That $T_{th}$ is different from $T_{FD}$ in this model thus means that the temperature does not reflect only the average energy, but also the amplitude of the energy fluctuations. For instance, as $T_{th} = \varepsilon/\eta$, a large value of $\eta$ implies a low temperature and correspond

Up to now, we have considered only the subclass ${\cal C}_s$ where $\psi(q)$ is a beta distribution, for which a form of DB holds. But what happens for more general distributions? In particular, one could wonder whether versions of the model with beta distributions are in some sense representative of the generic behavior of all the models belonging to $\cal C$. If $\psi(q)$ is different from a beta law, no DB holds \footnote{No simple form of DB seems to hold in this case, and  numerical simulations show the existence of local probability fluxes, thus confirming the absence of DB. Yet, we were not able to find an analytical proof for this result.}, and there is no clear way to find analytically the steady-state distribution $P_{st}(\{x_i\})$. Yet, numerical simulations show that even for distributions $\psi(q)$ far from beta laws, the two-point spatial correlation functions still vanish in steady state. This is also consistent with calculations made in the `q-model' for static sand piles, which exhibits som

\begin{figure}[t]
\centering\includegraphics[width=7.5cm,clip]{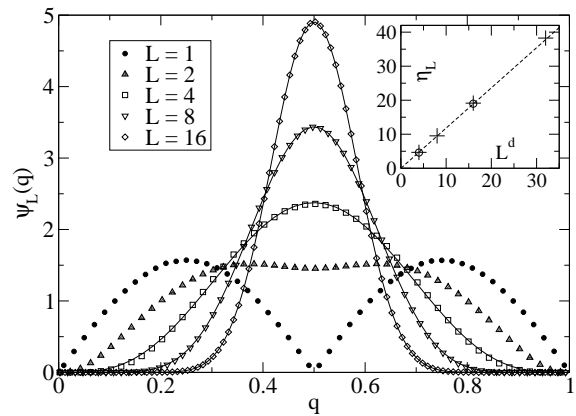}
\caption{\sl Renormalized distribution $\psi_L(q)$ for increasing sizes $L$, in dimension $d=1$. Full lines correspond to one-parameter fits with beta distributions. Inset: parameter $\eta_L$ from the fit vs.~$L^d$ for $d=1$ ($+$) and $d=2$ ($\circ$); dashed line is the mean-field prediction Eq.~(\protect{\ref{eq-eta_e}}).
}
\end{figure}

A standard way to coarse-grain the system is to introduce a renormalization group procedure for $\psi(q)$.
Since the distribution $P_{st}(\{x_i\})$ is not known, an analytical approach seems doomed from the outset. We thus implement numerically a functional renormalization group procedure in the following way: a large system is divided into blocks of $L^d$ sites each. Then one runs the dynamics and chooses a link in order to redistribute the energy. If both sites of the link belong to different blocks, then the energies $E_b^1$ and $E_b^2$ of the two blocks are computed, leading to the renormalized value $q_R = E_b^1/(E_b^1+E_b^2)$. The histogram of the values of $q_R$ obtained over the dynamics is recorded, yielding the renormalized distribution $\psi_L(q)$. If the initial distribution
$\psi(q)$ is a beta distribution, this renormalization transformation is exact (see below).
This procedure has been used for many  different initial distributions and
all of them lead to the same coarsed-grained result. This is illustrated 
on Fig.~1 for $L=2^n$ (data obtained by iterating $n$ times the renormalization with $L=2$), starting from the distribution $\psi(q)= \frac{\pi}{2}|\sin(2\pi q)|$ (filled circles). Very interestingly, $\psi_L(q)$
converges asymptotically when $L$ increases towards beta distributions with exponent $\eta_L$ linear in $L^d$ (Fig. 1). We have checked that the values of $q_R$ are decorrelated.

Starting with a beta distribution, $\eta_L$ can be computed exactly within our renormalization scheme. Indeed, the (one-site) distribution of the local energy $\varepsilon_i \equiv \frac{1}{2} x_i^2$ is a gamma distribution with parameter $\eta$. Since the $\varepsilon_i$'s are independent variables, the block energy $E_b$ also follows a gamma distribution of parameter $\eta_L=\eta L^d$, leading to a beta law with parameter $\eta_L$ for $q_R$. The important result here is that beta distributions are recovered numerically when starting from an arbitrary $\psi(q)$, even for moderate values of $L$ (Fig.~1). Consequently, an effective value of $\eta$ can be defined for any $\psi(q)$ as $\eta_e = \eta_L/L^d$ ($L \gg 1$); $\eta_e$ is the value of $\eta$ which would give the same coarse-grained behavior by using a beta $\psi(q)$ with parameter $\eta_e$ in the basic kinetic rules. Interestingly, $\eta_e$ can be computed within a mean-field approximation \cite{long}, yielding
\be \label{eq-eta_e}
\eta_e = \frac{1}{8 {\rm Var}(q)} - \frac{1}{2}
\ee
where ${\rm Var}(q) \equiv \langle q^2 \rangle - \langle q \rangle^2$ is the variance of the distribution $\psi(q)$. This value is in excellent agreement with the numerical simulations (Fig.~1).
The above results suggest an appealing scenario for the description of nonequilibrium systems with a conserved quantity and short-range correlations. Breaking the time-reversal symmetry leaves considerable freedom to choose the dynamics, but the renormalization group procedure shows that the macroscopic behavior can be described by a single parameter $\eta_e$ in addition to $T_{th}$. This new parameter $\eta_e$ essentially describes the way a globally conserved quantity is distributed among the different degrees of freedom. Its value is fixed in equilibrium: $\eta_{eq}=\frac{1}{2}$ here, but different values could be expected in other models --for instance, $\eta_{eq}=\frac{1}{p}$ for $E = \frac{1}{p} \sum_i |x_i|^p$. Note that the present approach uses a renormalization procedure in a context where there is a priori no diverging length scale in the system, i.e.~not close to a critical point \cite{Tauber}.

In conclusion, we have shown how to define a meaningful temperature $T_{th}$ from the conditional energy distribution of subsystems, within a class of finite-dimensional nonequilibrium models with conserved energy. The stationary distribution is generally non uniform over the states with given energy. The temperature deduced from FDR does not coincide with $T_{th}$, thus showing that FDR are not necessarily the most relevant way to define a temperature in out-of-equilibrium (and non glassy) steady-state systems. Finally, a numerical renormalization group approach indicates that DB is restored on a coarse-grained level even when this property is not satisfied microscopically. This renormalization procedure allows to define a parameter $\eta_e$ which encodes the deviation from equilibrium. The macroscopic behavior of the model is then described by the two parameters $T_{th}$ and $\eta_e$, i.e.~one more parameter than in equilibrium is required.

We thank J.-P.~Bouchaud and F.~Lequeux for important contributions, as well as P.~Sollich, J.~Kurchan and D.~Jou for interesting discussions.

\end{document}